\documentclass[11pt,twoside]{article}
\usepackage{asp2010}

\usepackage{color}
\usepackage{hyperref}
\hypersetup{colorlinks,citecolor=blue}

\resetcounters

\begin{document}

\title{WISE Discovery of Hyper Luminous Galaxies at z=2-4 and Their Implications for Galaxy and AGN Evolution$^\dagger$}\let\thefootnote\relax\footnote{$^{\dagger}$This publication makes use of data products from the {\it Wide-field Infrared Survey Explorer}, which is a joint project of the University of California, Los Angeles, and the Jet Propulsion Laboratory, California Institute of Technology, funded by the National Aeronautics and Space Administration.}
\author{Chao-Wei Tsai$^1$, Peter Eisenhardt$^2$, Jingwen Wu$^2$, Carrie Bridge$^3$, Roberto Assef$^2$, Dominic Benford$^4$, Andrew Blain$^5$, Roc Cutri$^1$, Roger L. Griffith$^1$, Thomas Jarrett$^1$, Carol Lonsdale$^6$, Sara Petty$^7$, Jack Sayers$^3$, Adam Stanford$^8$, Daniel Stern$^2$, Edward L. Wright$^7$, Lin Yan$^1$
\affil{$^1$Infrared Process and Analysis Center, California Institute of Technology}
\affil{$^2$Jet Propulsion Laboratory, California Institute of Technology}
\affil{$^3$Division of Physics, Math and Astronomy, California Institute of Technology}
\affil{$^4$Goddard Space Flight Center, NASA}
\affil{$^5$Department of Physics and Astronomy, University of Leicester}
\affil{$^6$National Radio Astronomy Observatory}
\affil{$^7$Department of Physics and Astronomy, UCLA}
\affil{$^8$Department of Physics, UC Davis}
}

\begin{abstract}
On behalf of the WISE Science team, we present the discovery of a class of distant dust-enshrouded galaxies with extremely high luminosity. These galaxies are selected to have extreme red colors in the mid-IR using NASA's Wide-field Infrared Survey Explorer (WISE). They are faint in the optical and near-IR, predominantly at z=2-4, and with IR luminosity $> 10^{13} L_{\sun}$, making them Hyper-Luminous Infrared Galaxies (HyLIRGs). SEDs incorporating the WISE, Spitzer, and Herschel PACS and SPIRE photometry indicate hot dust dominates the bolometric luminosity, presumably powered by AGN. Preliminary multi-wavelength follow-up suggests that they are different from normal populations in the local M-sigma relation. Their low source density implies that these objects are either intrinsically rare, or a short-lived phase in a more numerous population. If the latter is the case, these hot, dust-enshrouded galaxies may be an early stage in the interplay between AGN and galaxies.
\end{abstract}

\vspace{-5 mm}
\section{Introduction}
The NASA Wide-field Infrared Survey Explorer (WISE; PI: E.~L. Wright) launched on 2009 December 14, and completed its first full coverage of the sky in July 2010.  WISE achieves 5 $\sigma$ point source sensitivities of better than 0.08, 0.11, 1, and 6 mJy at 3.4, 4.6, 12 and 22 $\micron$ (hereafter W1, W2, W3, and W4) in a single coverage, consisting of 8 or more exposures at each sky location (Wright et al. 2010). 

One of the primary science goals of WISE is to identify extreme ULIRGs. At the longest 22 $\micron$ band, 
WISE is sensitive to the abundant hot dust radiation from the most powerful sources of mid-IR 
luminosity: both star-forming galaxies and AGNs. The WISE all-sky atlas is expected to include 
the most luminous galaxies in the Universe out to redshifts $\sim$ 3, reaching the epoch when major 
galaxy formation and assembly was underway (see recent review of Shapley 2011). These extreme objects may have been missed in smaller 
area surveys, and are far too dusty, and thus too faint for optical surveys (e.g. SDSS and 2dF). 

We have discovered a class of distant dust-enshrouded galaxies with extremely high luminosities. In this proceeding, we present a brief summary of current results of a multi-wavelength follow-up study on this population. More detailed discussions are presented in papers by Eisenhardt et al. (2012) and Wu et al. (2012). 

\vspace{-3 mm}
\section{Hunting for the HyLIRGs -- W12drops}
One of the approaches of searching for HyLIRGs which proved to be very successful is the so-called ``W1 W2 dropout'' selection. These galaxies, or W12 drops, are selected to have extreme red colors in the mid-IR (see Figure 2-left), making them faint or undetected in W1 and W2 and well-detected in W3 or W4, with W1 fluxes $<$ 0.034 mJy; and W4 $>$ 6.9 mJy and W4/W2 $>$ 92.5; or W3 $>$ 1.7 mJy and W3/W2 $>$ 22.4 (Eisenhardt et al 2012). As of early November of 2011, 75 redshifts extending up to z = 4.6 have been measured for W12 drops (primarily using Keck and Gemini-South), with 54 of them at z $>$ 1.6. The optical spectra reveal AGN activity in $\sim$ 50\% of W12drops.

Our follow-up programs to determine the total far-IR luminosity ($L_{IR}$) for high-z W12 drops reveal their $L_{IR}$ is consistently $> 10^{13} L_{\sun}$, putting them in the category of HyLIRGs. Preliminary adaptive optics imaging results on W12drops show no obvious lensing features. We estimate that there are a total of $\sim$ 1,000 hyper luminous W12 drops over the sky, making them unlikely to be covered by most deep surveys.

\subsection{Ongoing Multi-wavelength Followup Programs}
The members on the team are working to establish the comprehensive SEDs of W12 drops. At optical and near-IR wavelengths, we are using the Palomar 200in and NOAO 4m telescopes for imaging, and Keck, Gemini-S, Magellan, and MMT for spectroscopy. Our HST programs aim to probe the stellar populations, search for possible lensing features, and investigate the environments of the W12 Drops. In the mid-IR, in addition to WISE data, our Spitzer programs provide deep 3-5 \micron\ photometry well below the WISE limits. In the submm region, the  Herschel provides 70 to 500 micron photometry, giving reliable measurements of $L_{FIR}$ and dust temperatures in our systems. We also have intensive millimeter and submillimeter measurement programs with CSO, JCMT, IRAM30m, and CARMA underway which provide information on colder dust components in W12 drops (Wu, et al. 2012). The ATCA and EVLA programs are aiming to measure radio continuum emission from two selected W12 drop targets. The X-ray activity of the obscured AGNs will be studied by XMM and NuSTAR.

\subsection{WISE 1814+3412 -- First HyLIRG Discovered by WISE}
As described in Eisenhardt et al. (2012), WISE 1814+3412 was first identified in 2010 May using the W1W2 dropout criteria, and spectroscopically confirmed to be at z=2.452 using LRIS on Keck I. The LRIS spectrum is typical of a Lyman Break Galaxy (LBG) in the redshift range without AGN features. The rest frame UV spectrum suggests the star formation rate (SFR) $> 100\,M_{\sun}\,yr^{-1}$. The SED can be decomposed into three components: disk, starburst, and obscured AGN (see Figure 1). The required high obscuration on the AGN component ($A_{V} \sim 50$) is consistent with the optical spectroscopy where the features of AGN activity are absent.

\begin{figure}
\plotone[height=5.4cm]{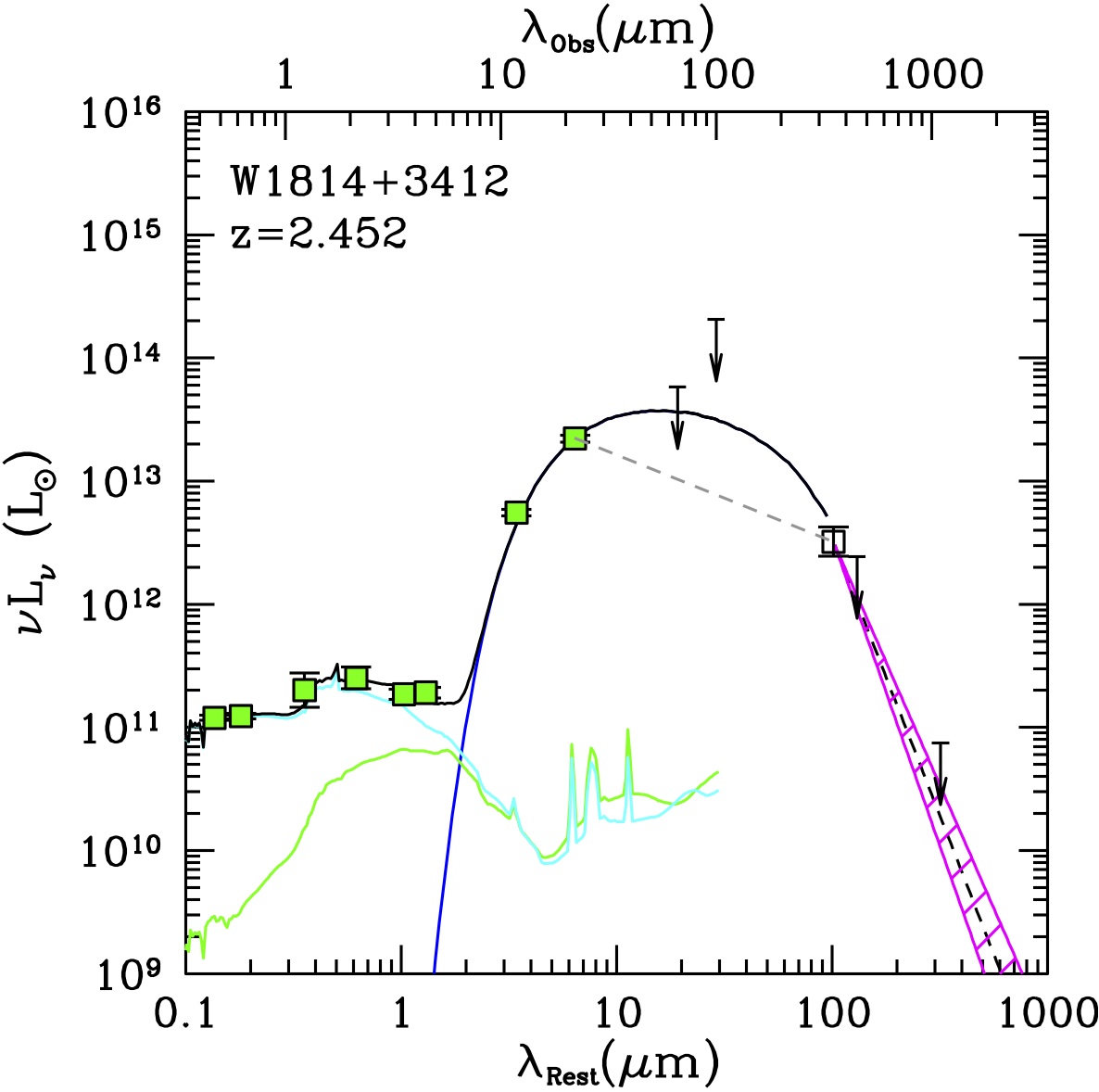}
\caption{Best fit SED model to the photometry 
(green symbols) for the z = 2.452 W12drop WISE 1814+3412 plotted in $\nu L_{\nu}$ units, reproduced from Eisenhardt et al (2012). 
The fit contains starburst (cyan), Sbc spiral (green), and AGN (blue) components, the latter with $A_{V} \sim 50$ extinction applied, and extended into the submm with $F_{\nu} \propto \nu^{3}$ to $\nu^{4}$ (magenta). No signiÞcant cold dust component is needed.}
\label{figure:W1814_SED}
\end{figure}

The minimum bolometric luminosity ($L_{bol}$) using a simple integration over the power-law interpolations between measured data points is $3.7 \times 10^{13} L_{\sun}$, and more likely to be $\sim 9 \times 10^{13} L_{\sun}$ (based on SED models). The flux ratio between W4 and 350 \micron\ indicates dust warmer than 150K dominates the luminosity. In contrast, the 350 \micron\ to 1.1 mm flux ratio suggests that dust cooler than 35K (typical dust temperature of sub-mm galaxies; Kovacs et al 2006) contributes insignificantly to the bolometric luminosity in this source. The $L_{bol}$ corresponding to the UV SFR of $> 100\,M_{\sun}\,yr^{-1}$ is less than 10\%\ of the total $L_{bol}$. Despite the relatively small contribution in luminosity, strong star formation and AGN activity are coexisting in this system.

Keck near IR images with Laser Guide Star Adaptive Optics show that WISE 1814+3412 has four well-resolved components in a 10\arcsec region, separated by $\sim$ 2\arcsec--3\arcsec\ from each other. The LRIS spectra reveal that three of them, including the primary source at the position of the mid-IR peak, are at the same redshift of 2.45, while the fourth component is a foreground star. Although the three components at the same redshift suggest gravitational lensing in this system, the three objects have quite different spectroscopic characteristics. This argues against the lensing possibility, and the high luminosity of WISE 1814+3412 is likely to be intrinsic. 

\subsection{Total Luminosity in Mid-IR to Submillimeter}

\begin{figure}
\plotone[height=4.6cm]{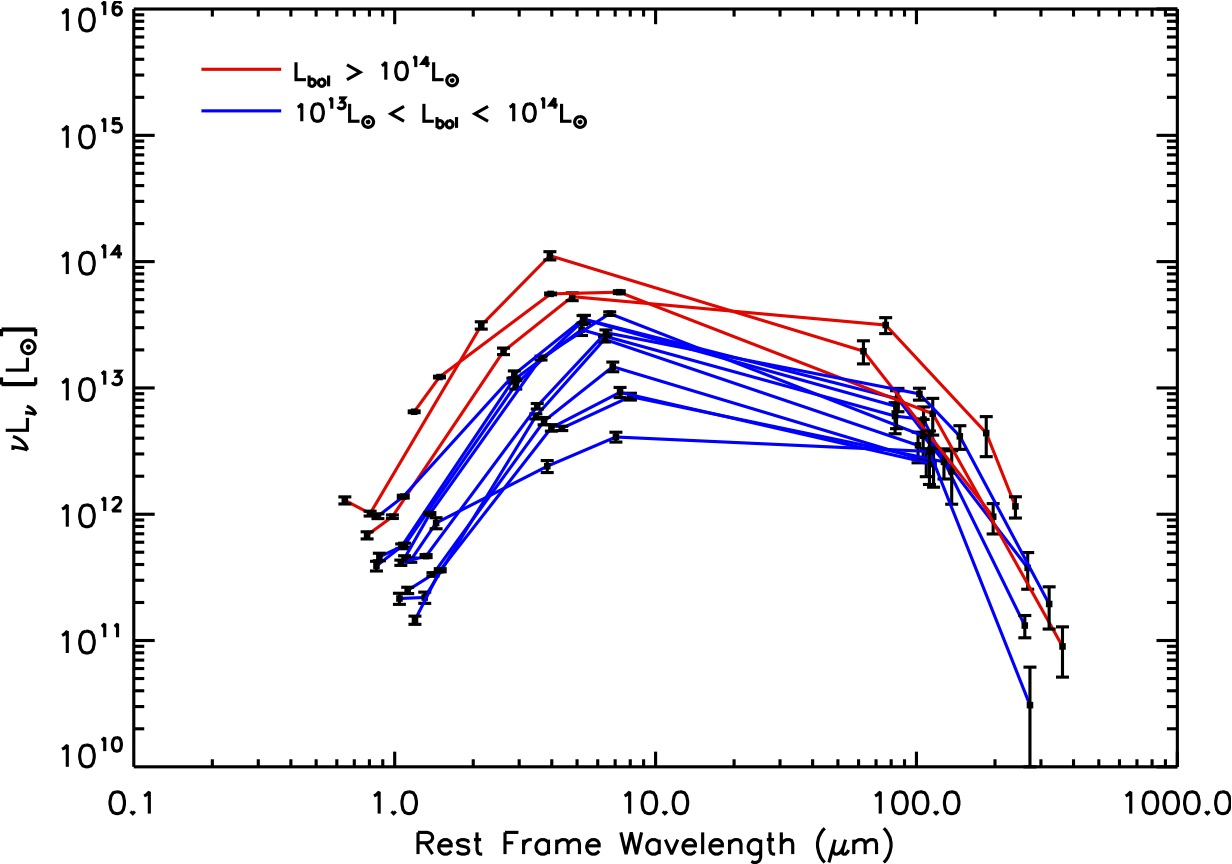}\plotone[height=4.6cm]{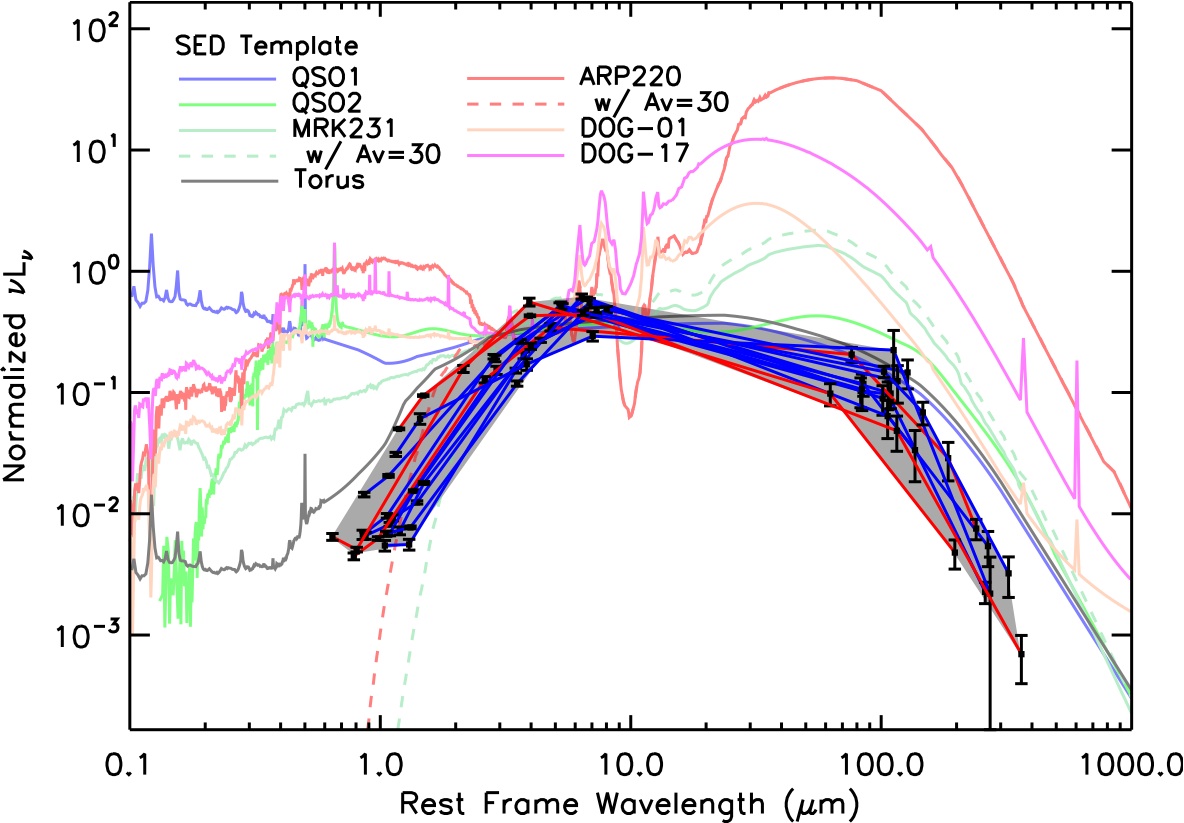}
\caption{\textit{left}: W12drop SEDs in the rest frame with submm detections from CSO in unit of $\nu L_{\nu}$. \textit{right}: $\nu L_{\nu}$ SEDs normalized to the $L_{bol}$. The grey region highlights the assembly of normalized SEDs, while the color curves show a range of standard ULIRG SED templates, scaled to 20\%\ of $L_{bol}$ at 5\micron.}
\label{figure:W12drops_SED}
\end{figure}

The SEDs of W12drops with ground-based submillimeter measurements (Wu et al. 2012) are shown in Figure 2 (\textit{left}), including Spitzer/IRAC data at the shortest wavelengths. The SEDs in general peak in the rest-frame mid-IR in $\nu L_{\nu}$, and are flat toward longer wavelengths until $\sim$ 100 \micron. This indicates that the majority of the luminosity of this type comes from the mid-IR, presumably very hot dust. The conservative estimates of total $L_{bol}$ are all above $10^{13} L_{\sun}$, with some reaching $10^{14} L_{\sun}$. Normalizing the SEDs by the $L_{bol}$ (\textit{right} panel of Figure 2) shows a tighter sequence, representing the typical SED of W12drops. This prototypical SED of W12 drops is quite different from other known luminous populations, being steeper at both ends of the SED. Simple dust temperature modeling provides a lower limit of the dust component of $>$ 35K. This simply means that the SEDs of W12 drops are overwhelmed by the mid-IR, hot dust ($T_{d} >$ 100K), and does not necessarily imply lack of cold dust ($T_{d} <$ 35K). Since the dust emission goes as $T_{d}^4$, hotter dust doesn't need to be abundant to dominate the $L_{bol}$.  

\vspace{-3 mm}
\section{Summary}
The newly discovered HyLIRG population ``W12 dropouts'' is presented. With $L_{bol} > 10^{13} L_{\sun}$, their SEDs  are dominated by a hot dust component, presumably powered by highly obscured AGNs. Preliminary results from high resolution imaging show no obvious lensing signatures, thus their high luminosity is likely to be intrinsic. In WISE 1814+3412, the first discovered WISE HyLIRG, the optical (rest frame UV) data show a high SFR in the system. This system is likely hosting ongoing starburst and strong AGN activity. 
We found $\sim$ 1,000 W12drops from the WISE whole sky survey The low surface density implies that they are either intrinsically rare, or are a short-lived phase in a larger population. It is possible that this W12drop population is in a transition phase when the AGN has just turned on, and the star formation in the host galaxy has not yet ceased.

\end{document}